\documentstyle[emulateapj,psfig]{article} 

\slugcomment{Accepted by the Astrophysical Journal}

\lefthead{Marr et al.}
\righthead{Non-Uniform Free-Free Absorption in 0108+388}

\begin{document}

\title{Non-Uniform Free-Free Absorption in the GPS Radio Galaxy 0108+388}

\author{J. M. Marr\altaffilmark{1}, G. B. Taylor\altaffilmark{2}, and 
F. Crawford\altaffilmark{3}}

\altaffiltext{1}{Union College, Schenectady, NY  12308, USA; marrj@union.edu}
\altaffiltext{2}{NRAO, Socorro, NM  87801, USA; gtaylor@nrao.edu}
\altaffiltext{3}{MIT, Cambridge, MA  02139, USA; crawford@space.mit.edu}

\begin{abstract}

We have observed the canonical gigahertz-peaked spectrum source 0108+388
with the VLBA at a range of frequencies above and below the spectral peak.
The activity that dominates the radio emission from 0108+388, which is
also classified as a Compact Symmetric Object, is thought to be less than
1000 years old.  We present strong evidence that the spectral turnover
in 0108+388 results from free-free absorption by non-uniform gas, possibly
in the form of a disk in the central tens of parsecs.  
\end{abstract}


\section{INTRODUCTION}

The radio source 0108+388 (RA=$01^{\rm h} 11^{\rm m} 37^{\rm s}.32$,
Dec=$39^\circ 06'28''.1$ (J2000), $z$=0.699) is an example of a class
of extragalactic radio source referred to as Gigahertz-Peaked Spectrum
(or GPS) sources.  GPS sources, which are discussed at length in a
thorough review by O'Dea (1998), are characterized by large radio
luminosities, compact structure, and spectra that turn over at
Gigahertz frequencies with steep spectra on either side of the peak
(O'Dea, Baum, \& Stanghellini 1991; O'Dea 1998).  The steep slope at
high frequencies is generally accepted to be due to synchrotron
losses, but there are two main competing theories for the cause of the
low frequency turnover.  Either the lower frequency radio emission is
absorbed by ionized gas in the line of sight in the host galaxies,
with conditions comparable to that of emission-line regions in active
galaxies (van Breugel, Heckman, \& Miley 1984, Bicknell, Dopita, \&
O'Dea 1997; Kameno et al.\ 2000), or the nuclei of these galaxies have
exceptionally large magnetic fields, causing synchrotron
self-absorption at GHz frequencies (Hodges, Mutel, \& Phillips 1984;
Mutel, Hodges, \& Phillips 1985; O'Dea et al.\ 1991; Readhead et al.\
1996; de Vries, Barthel, \& O'Dea 1997; Snellen et al.\ 2000).
Absorption by induced Compton scattering in combination with free-free
absorption has also been proposed by Kuncic, Bicknell, \& Dopita
(1998).

As GPS sources have been studied in more detail since the discovery of
a large population of such sources in the 1980's (Gopal-Krishna,
Patnaik, \& Steppe 1983, Spoelstra, Patnaik, \& Gopal-Krishna 1985),
it has become apparent that they constitute a diverse group, with
morphological and spectral variances.  The object we discuss here,
0108+388, is a canonical case of a morphological group referred to by
Wilkinson et al.\ (1994) as compact symmetric objects (or CSO's),
which have small ($<$ 1 kpc) symmetric radio emission structure.
Multi-epoch VLBI observations of a number of CSO's have revealed
expansion rates that imply ages of the order of 1000 years (Owsianik,
Conway, \& Polatidis 1998; Owsianik \& Conway 1998; Taylor et al.\
2000), supporting the earlier hypothesis that CSO's are precursors to
the young classical double radio galaxies (Phillips \& Mutel 1982;
Carvalho 1985; Readhead et al.\ 1996).  VLBI observations of 0108+388
by Owsianik et al.\ (1998) and by Taylor et al.\ (2000) infer apparent
separation speeds of the outer components of 0.197 $\pm$ 0.026
$h^{-1}$ c and 0.24 $\pm$ 0.04 $h^{-1}$ c, respectively, from which
these authors derive kinematic ages of just 367 $\pm$ 48 yr and 310
$\pm$ 70 yr, respectively.  This is the youngest age yet measured for
a radio galaxy.  However, with VLA and WSRT observations of 0108+388
Baum et al.\ (1990) detected extended emission 20 arcsec (78 $h^{-1}$
kpc) from the core, casting an apparent contradiction to the young
classical double radio galaxy model.  Baum et al.\ propose that either
the activity in 0108+388 is recurrent or has recently been smothered
by infalling gas.

High resolution imaging of 0108+388 was first undertaken at 5 GHz as
part of the PR VLBI survey (Pearson \& Readhead 1988) and found to
have a simple double structure.  Conway et al.\ (1994) and Taylor,
Readhead, \& Pearson (1996) made more sensitive and higher resolution
images at higher frequencies that revealed a steeply inverted core
component connected to both outer components by a faint chain of
components.  Overall, 0108+388 has an `S' symmetry that is relatively
common among CSOs (Readhead et al.\ 1996).
 
The spectrum of the total radio emission from the compact structure in
0108+388 was analyzed by Baum et al.\ (1990).  They measured a
spectral index below the turnover that was steep and roughly constant,
in conflict with the exponential turnover expected from free-free
absorption.  However, their observed spectrum is the total from
several components in the source.  An exponential turnover in the
free-free absorption case will apply only if the absorbing medium
uniformly covers the entire radio source.  We report here on a set of
VLBA observations designed to study the absorption at each point in
the map.  We have observed 0108+388 with the VLBA at five frequencies:
two above the turnover, two below the turnover, and one at the
turnover frequency itself.  The maps made with these observations
reveal the morphology of the radio emission from 0108+388 and how it
is obscured by absorption.  We find that the spectra of 0108+388 at
all positions in our maps are consistent with free-free absorption.
Additionally, our maps reveal that the absorption is non-uniform and
is suggestive of an edge-on disk centered on the core.  Other evidence
for a large amount of thermal gas along the line of sight to the core
of 0108+388 has been reported by Carilli et al.\ (1998), who measured
significant absorption by H{\sc i}, and by Cawthorne et al.\ (1993),
who find no measurable polarization of the radio emission. The
existence of a circumnuclear gas disk, with likely instabilities,
could also be the source of intermittent fueling of the central engine
and explain the recent onset of activity along with the past activity.

\section{OBSERVATIONS AND MAPPING}

We observed 0108+388 with the Very Long Baseline Array
(VLBA)\footnote{The VLBA is operated by the National Radio Astronomy
Observatory, which is a facility of the National Science Foundation
operated under cooperative agreement by Associated Universities, Inc.}
at 8.417, 4.983, 2.267, and 1.663 GHz on 1996 June 24 and at 15.359
and 8.415 GHz on 1997 February 03.  In the 1996 observations, the
source signal was recorded at each station for a total of 100 to 112.5
minutes at each frequency, with an integration time of 2 seconds.  In
the 1997 observations a total of 330 minutes on source was obtained at
both 8.415 and 15.359 GHz.  The on-source scans at all frequencies and
both epochs were spread across a wide range in hour angle in order to
obtain good $u$-$v$ coverage.

Fringe fitting was performed with AIPS using the small and bright
source J0136+478 as a fringe calibrator.  In the fringe searches we
generally used solution intervals of 3 minutes, moderately large delay
and rate windows (100 ns and 40 mHz), and a minimum signal-to-noise
ratio of 7.  Good solutions were generally found at small and stable
delays and rates.  We applied {\it a priori} amplitude calibration,
also with AIPS, using the system temperatures and gains measured by
the NRAO staff.  We assume that the amplitude calibration is accurate
to within 5\%.  The detected flux densities on the shortest baselines
were all greater than 92\% of previously reported flux densities from
low resolution observations (Baum et al.\ 1990; O'Dea et al.\ 1990;
UMRAO database).

Editing, modelfitting, and mapping were done using Difmap (Shepherd,
Pearson, \& Taylor 1994, 1995; Shepherd 1997).  Starting with the 1997
15-GHz data, we used an initial model of seven elliptical gaussian
components, based on the appearance of the clean map made by Taylor et
al.\ (1996).  The model parameters were then allowed to vary to yield
a best rms-fit in the $u$-$v$ plane to the self-calibrated data.  We
found best-fit models for each data set independently except that we
checked for general consistency in the positions of components at all
frequencies.  The final model for each data set fit the
self-calibrated data with an rms difference of around $1\sigma$.

We fed the best-fit models as input models for Hybrid CLEAN (Readhead
\& Wilkinson 1978), which we used to obtain our final image maps.  We
initially used small CLEAN boxes that barely encompassed the apparent
emission in the dirty map.  After 100 CLEAN components were selected,
with a gain of just 0.05, we increased the sizes of the CLEAN boxes
while always staying marginally larger than the area of the apparent
flux in the residual map.  We excluded apparent spurious features from
the CLEAN boxes the first few cycles of self-cal and CLEAN.  CLEANing
was generally stopped a few clean components before the first negative
component.  The maps were of sufficiently high quality that all the
detected flux on the shortest baselines was collected in the CLEANing
by this point.  We did not apply amplitude self-calibration until the
model agreed with the data with an rms difference less than
$1.5\sigma.$ The final clean component models fit the data to better
than $1.0\sigma.$ The noise in the maps range from 0.2 mJy/beam in the
8.4-GHz maps to 0.7 mJy/beam in the low frequency maps, yielding
signal-to-noise ratios from 300 to over 1000.

We also made spectral index maps between neighboring frequency bands
using the AIPS task COMB.  We first edited the data at each frequency,
removing the larger $u$-$v$ spacings from the higher frequency data
and the smaller $u$-$v$ spacings from the lower frequency data, and
remapped in Difmap.  The maps at the two frequencies were convolved
with identical clean beams.  In this way, we produced spectral index
maps between 1.7 and 2.3 GHz, between 2.3 and 5.0 GHz, between 5.0 and
8.4 GHz from the 1996 June observations, and between 8.4 and 15.4 GHz
from the 1997 February observations.  Since moderate percentage errors
in the observed maps propagate to large errors in the spectral index,
we masked the spectral index maps where the signal in either clean map
was less than 10$\sigma$.

Since the observations were not phase-referenced to a nearby
calibrator, no absolute position information is available for each
map.  Each map, therefore, can be arbitrarily shifted in RA and Dec,
and so the correct alignment of the maps is not known {\it a priori}.
This is a critical issue in making spectral index maps.  If the maps
are misaligned, extremely large and extremely small spectral indices
will be inferred, particularly at the edges of the emission regions.
At the higher frequencies (from 5.0 to 15.4 GHz) the relative
positions of the component peaks were consistent such that they could
be used to determine the alignments of the maps.  The resultant
spectral index maps were quite reasonable in that they did not show
surprisingly large spectral indices at the edges or sides of the
features.  At the lower frequencies, however, the separations of the
centroids of the components varied.  For these spectral index maps, we
chose alignments that yielded the minimum magnitudes of the extreme
spectral indices.  Additionally, since the flux-density distribution
in the maps appears relatively symmetric, especially in the lower
resolution maps, we also checked the alignment of the maps by
examining the spectral index map and watching for dipolar
distributions of inferred spectral indices.  We successfully found
relative alignments that minimized the spectral index extremes and
that yielded roughly symmetrical spectral indices.

\section{RESULTS}

The CLEAN maps for each observation are shown in Figure 1.  As noted
by previous investigators of this source (Pearson \& Readhead 1988;
Conway et al.\ 1994; Taylor et al.\ 1996; Owsianik \& Conway 1998;
Owsianik et al.\ 1998) the intensity distribution of the source at all
frequencies consistently involves two major components.  Following the
notation of Taylor et al.\ (1996), we refer to these components as C1
(in the southwest) and C7 (in the northeast).  At the higher
frequencies (15.4 and 8.4 GHz) we obtained models with a total of
seven components, while at the three lower frequencies we could
reasonably fit only four components, due to blending of the smaller
components.

\subsection{Spectral Index Maps}

The spectral index maps are shown in Figure 2.  The uncertainties in
the inferred spectral indices, calculated by propagating both
uncertainties in amplitude calibration and the map noise, are listed
in Table 1.

The spectral index map between the two highest frequencies shows that
the two major components have steeply declining spectra, with spectral
indices $\alpha \approx -1.0$, where $F_\nu \propto \nu^\alpha$.  A
small region between the two major components, centered on component
C3, in the notation of Taylor et al.\ (1996), has an inverted spectrum
with $\alpha$ getting as large as +1.8, suggesting that the core of
this active galaxy is located here, as previously noted by Taylor et
al.\ (1996).

Between 8.4 and 5.0 GHz, the spectra of the regions containing the
major components are generally flatter than between 15.4 and 8.4 GHz,
although the spectra get steeper towards the outer edges.  At 5.0 GHz
the core region has small flux densities, so most of this central
region is masked in the spectral index map.

Between 5.0 and 2.3 GHz, the spectral index map shows that the spectra
of all but the easternmost and westernmost edges are inverted ($\alpha
> 0$).  At the very edges of the eastern and western extremes the
spectral indices are still seen to get as low as $-$0.7.  The region
with inverted spectra appears to run across the map in the north-south
direction, but with the steepest parts located in two areas north and
south of the core at the edge of the mapped region.  The core is not
detected at these frequencies due to its low luminosities and steeply
inverted spectrum.

Between 2.3 and 1.7 GHz, the spectral index map looks very similar to
the 2.3-5.0 GHz spectral index map, except that all the spectral
indices across the entire map are larger and the region of inverted
spectra encompasses almost the entire source.  The two areas of the
steepest inverted spectra are more evident and appear even more
symmetrical about the core.

\subsection{Free-free Absorption Model}

As clearly shown in Figure 2 the turnover of the composite spectrum of
0108+388 at 5.0 GHz results from the majority of the source, including
the two brightest components of the source, turning over at this
frequency.  This is in conflict with the expectations for synchrotron
self-absorption, in which the small regions with the largest magnetic
fields and electron densities would turn over at higher frequencies.
The nature of the turnover in 0108+388, however, is consistent with
the absorption being due to free-free absorption by a foreground
screen of ionized gas.  Additionally, the peak spectral index between
2.3 and 1.7 GHz, even at the moderately low resolution of the map in
Figure 2d, is 3.2 $\pm$ 0.5.  This is larger than the maximum spectral
index allowed by classical self-absorption models, although it can be
fit to synchrotron self-absorption with special physical conditions
(see, e.g., de Kool \& Begelman 1989).

The spectral index maps also suggest the shape and structure of the
absorbing gas, which appears to be extended north-south with two
density peaks on either side of the center.  This is suggestive of an
edge-on disk centered on the core of the galaxy.  However, the
spectral index only measures the ratio of the flux at two frequencies
and does not directly indicate the densities of the absorbing gas.  To
directly map the absorbing gas we made opacity maps, which we discuss
in the following section.

\subsection{Opacity Maps}

The creation of opacity maps involves comparison of maps at all the
observed frequencies (as described below).  For this purpose, all data
sets were re-mapped with the same resolution.  Two methods were used.
First, we simply convolved all clean maps with the same beam as the
clean beam at 1.7 GHz.  This approach is flawed, in general, because
the different frequency observations (with the same array) probe
structures of different sizes, and so even with the same clean beam
different features are being compared.  As a second approach we edited
all data sets to have the same minimum and maximum $u$-$v$ distances;
all data sets were flagged to have the same maximum $u$-$v$ distance
as in the 1.7-GHz data set and the same minimum $u$-$v$ distance as in
the 15.4-GHz data.  These truncated data sets had aspect ratios of
only 9.0 and so this method could result in the maps missing much of
the structure, both on large and on small scales.  For example, the
detected flux densities on the remaining shortest baselines at 2.7 and
1.3 GHz were 87\% and 78\%, respectively, of those in the full data
sets.  Interestingly, there were no significant differences in the
final maps made with these two methods and the maps looked very much
like the 2.3 and 1.7 GHz clean maps in Figure 1.  Furthermore the
signal to noise ratios in the final maps made from the edited data
sets were, in general, higher than in the clean maps from the
untruncated data sets, especially at the mid-frequencies; this
resulted primarily because the editing removed the data at the
extremes in the $u$-$v$ plane, which corresponds to where the density
of data is lowest.  The high fidelity of these maps resulted from the
simplicity of the structure of this source at these low frequencies.
The clean maps made from the truncated data sets are shown in Figure
3.

To infer optical depths we need flux-density maps in the absence of
the absorption.  We obtain such maps by extrapolating the spectra at
the higher frequencies.  Since the spectra of the major components
between 15.4 and 8.4 GHz are steeply declining with increasing
frequency, the absorption at these frequencies is small and so the
flux densities at these frequencies provide a close approximation of
the unabsorbed emission spectrum.  If $\alpha_{15-8}(x,y)$ is the
spectral index between 15.4 and 8.4 GHz at some position $(x,y)$ in the
map, and $F_{\rm 8.4GHz}(x,y)$ is the flux density at 8.4 GHz at this
position, then the approximate unabsorbed flux density at 2.3 GHz is
$${\rm Unabsorbed}~F_{\rm 2.3GHz}(x,y) = F_{\rm 8.4GHz}(x,y)
\left[ {{\rm 2.3~GHz} \over {\rm 8.4~GHz}}\right] ^{\alpha_{15-8}(x,y)}.\eqno(1)$$  
We then obtain an approximation
of the optical depth at 2.3 GHz at each point in the map by
$$\tau(x,y)=-\ln({\rm observed}~F_\nu(x,y)) + \ln({\rm
unabsorbed}~F_\nu(x,y)).\eqno(2)$$ However, since the optical depths
at 8.4 and 15.4 GHz are not really zero, the extrapolation of the
spectrum produces a small error.  We improve the inferred optical
depths by iterating the procedure; we use the inferred optical depths
at 2.3 GHz to infer the optical depths at 8.4 and 15.4 GHz, assuming
free-free absorption.  We then correct the 8.4 and 15.4 GHz maps for
the non-zero optical depths, yielding maps of the unabsorbed
flux-densities at these frequencies.  We then used these modified maps
to recreate the 15.4-8.4 GHz spectral index map and the unabsorbed 1.7
and 2.3 GHz flux density maps, which we used, in turn, to create
improved opacity maps.

The opacity maps were made from the maps shown in Figure 3 by using a
number of steps of the AIPS task COMB which reproduced the math
discussed in the previous paragraph.  The low-frequency flux densities
(at 5.0, 2.3, and 1.7 GHz) were extrapolated from the 8.4-GHz flux
densities of the same epoch (1996 June) while the spectral indices
between 15.4 GHz and 8.4 GHz ($\alpha_{15-8}(x,y)$) were calculated
using the 8.4-GHz data of the same epoch (1997 February) as the
15.4-GHz data.  The resultant opacity maps at 2.3 and 1.7 GHz are
shown in Figure 4.  The uncertainties in the optical depths are
primarily due to the 5\% uncertainty in amplitude calibration, which
yields a constant uncertainty across each map.  The calculated
uncertainties due to amplitude calibration errors are listed in Table
2.  The opacity maps at 5.0, 2.3, and 1.7 GHz are nearly identical,
except that the optical depths are larger at lower frequencies, as
expected.  The optical depths at 5.0 GHz are small, relative to the
uncertainty, and so the 5.0-GHz opacity map is dominated by noise and
so is not displayed in Figure 4.  The opacity maps are nearly
identical in appearance.  These maps also indeed show that the maximum
absorption occurs at two points north and south of the center and that
the optical depth decreases in the direction perpendicular to the line
connecting the optical depth peaks.

At each point in the map, we have a measure of the optical depth at
three different frequencies, and so we can check whether free-free absorption
fits the observed spectra everywhere.  The optical depth of free-free
absorption, in a simple single-component model, depends on frequency 
according to
$$\tau_\nu = 0.08235 ~\nu^{-2.1} \int T_e^{-1.35} N_e^2 dl,$$ where
$\nu < 10$ GHz and is given in GHz, $N_e$ is in units of cm$^{-3}$,
and $L$ is in pc (Mezger \& Henderson 1967).  Assuming this model, we
can remove the frequency dependence in our opacity maps, and check for
consistency between the data at different frequencies, by multiplying
the optical depths in the maps of Figure 4 by $\nu^{2.1}$ (using the
AIPS task COMB).  If each position in these maps has a single
free-free absorption component, then the resultant images will display
maps of the value $0.08235 \int T_e^{-1.35} N_e^2 dl$.  These maps are
shown in Figure 5 (again, because of the large uncertainty relative to
calculated values we do not display the map at 5.0 GHz).  The
uncertainties in these values, calculated by propagating through just
the amplitude calibration errors, are listed in Table 2 (the total
uncertainties actually include errors due to map noise, which are
significant only near the edges of the source).  The maps in Figure 5
are essentially identical.  In Figure 6 we display the difference
between the two maps in Figure 5 divided by the total uncertainty, due
to both amplitude calibration errors and map noise.  The contours
displayed in Figure 6 represent difference values of $-$0.5, 0.5, 1,
and 2 times the calculated uncertainty.  The differences are typically
$\approx 1.0\sigma$ across the map, with a mean of 1.1$\sigma$, and
get as large as 2.1$\sigma$ only near the southwest edge.  The rms
value in Figure 6, and hence the rms deviation between the maps in
Figure 5, is 1.26$\sigma$.

As is evident in Figures 4 and 5, the optical depth is not uniform
over the map and most likely varies on scales smaller than the
resolution of these maps.  Therefore, one should not expect the
single-component free-free absorption model to perfectly fit the
spectra at each point in our moderately low resolution maps.
Therefore, a reduced $\chi^2$ value for the agreement between this
simple model and our data across our entire map of just 1.26 is
striking and strongly supports the premise that free-free absorption
is at work.  Only a mild level of complexity, such as a variable
electron density across our synthesized beam, would be needed to
produce a fit with an rms deviation of less than 1$\sigma$.

\subsection{Individual Spectra}

As an additional test, one can fit a free-free absorption model to
individual spectra.  By stacking the maps in Figure 3, with the same
alignments that we used to make the opacity maps, we can obtain
individual spectra at each location.  However, as we discussed in the
previous section, one should expect that the optical depth varies on
scales below the low resolution of the maps in Figure 3.  One should
not, therefore, expect the precise fitting of a single-component
free-free absorption model to the observed spectra to produce a
perfect fit.

In Figure 7, we display the spectra at the locations of the peaks in
emission at 5.0 GHz.  To correct for any flux variations in the source
between our two observation epochs the flux densities at 15.4 GHz in
our 1997 observation were multiplied by a scale factor of 0.97, which
is the ratio of the 8.4-GHz flux densities in 1996 to that in 1997.
The displayed error bars correspond to the total uncertainty in the
measured flux densities, due to both the 5\% amplitude calibration
uncertainty and the rms in the map flux densities.

Overlaid on the observed spectra in Figure 7 are best-fit curves for
single-component models of free-free absorption in the foreground of
optically thin synchrotron emission sources.  The curves in Figure 4
are given by $F_\nu = F_o (\nu / 1.5359 {\rm GHz})^\alpha \exp
(-EMT/\nu^{2.1})$. Since our goal is to determine the cause of the
absorption at low frequencies, we kept $F_o$ relatively fixed to
ensure that the curve passes through the measured 15.4-GHz flux
density.  Variation in $EMT$, though, causes slight changes in the model
15.4-GHz flux density, so we needed to allow $F_o$ to vary a little,
as $EMT$ varied, to ensure that the flux-density at 15.4 GHz matched the
observed value.  We, therefore, had two free parameters.  The values
of these parameters in the best fit curves are listed in Table 3.
Although the curves appear by visual inspection to fit the data well,
the reduced $\chi^2$ of the fits, listed in Table 3, are quite large.
Almost the entire contribution to these large $\chi^2$ values are in
the disagreements at 2.3 and 1.7 GHz, where the slopes of the curves
are very steep and small adjustments in the model parameters yield
large differences in the curve values.  However, a slightly more
complex model, such as an electron density that varies by a factor of
two on scales below the map resolution, yields nearly perfect fits to
the data.  We, therefore, find that the observed spectra at these two
positions are consistent with a reasonable free-free absorption model.

\section{DISCUSSION}

\subsection{Disk of Ionized Gas}

The maps in Figure 5, which show the morphology of the quantity
$0.08235 \int T_e^{-1.35} N_e^2 dl$ projected onto the sky plane, show
that the absorbing medium is elongated north-south and is concentrated
in two spots symmetrical about the core.  In Figure 8 we display the
1.7-GHz map from Figure 5 in contours overlaid on a grey-scale plot of
the 15.4-GHz clean map from Figure 3.  Even with the large clean beam
of these maps, the denser regions of the absorbing medium are clearly
shown to be located to the sides of the radio emission structure.
Such a morphology in optical depth is often symptomatic of a disk
centered on the core.  If this is indeed a disk, it is roughly
perpendicular to the jet axis.  One must keep in mind that this medium
is probed in our maps only through its absorption of the radio
emission from the jet, and so the medium can, and most likely does,
extend well beyond the regions indicated in our maps.  Additionally,
the absorption is caused only by ionized gas and any such disk is most
likely to have a large neutral component that is far more extensive.
In fact, the observations of Carilli et al.\ (1998) indicate that a
large amount of neutral gas does in fact exist in this galaxy.

In interpreting the morphology of the opacity maps, one must keep in
mind that these maps were made with a resolution of 4.4 mas in the
North-South direction and 3 mas in the East-West direction.  For
example, the peaks in optical depth in Figure 4 appear to be located
where there is no detectable emission at 15.4 GHz in the high
resolution map of Figure 1a.  However, since in the process with which
we made the opacity maps the input maps were masked where the signal
was below 10$\sigma$, this doesn't seem possible.  The reason that we
appear to detect absorption far from the radio emission structure seen
in Figure 1a is because of the low resolution of the opacity maps; the
15.4-GHz map of Figure 3a was used in making the opacity maps, not
Figure 1a.  If we were able to make opacity maps at the resolution of
Figure 1a, we would find the optical depth peaks to occur somewhere at
the edges of the 15.4-GHz emission.  In the opacity maps of Figure 4,
the precision of the position information is given by the width of the
clean beam, and so a true location of the optical depth peaks 1 mas
away from the apparent location is reasonable.  The morphological
information in our opacity maps of which we can be sure is that
indicated in Figure 8, i.e. that the peaks in the optical depth are
offset from the peaks in the radio emission.  In order for the optical
depth peaks to appear at the edges of our opacity maps, the true
optical depth structure must have a similar gradient.

The peaks in optical depth are located about 10 $\pm$ 4.4 mas apart.
Therefore, if this is indicative of a disk, then the disk appears to
have an inner radius of about 5 $\pm$ 2.2 mas.  With a redshift of
0.699 and a Hubble constant of $100~h~$km~s$^{-1}$ Mpc$^{-1}$, this
corresponds to an inner radius of $\approx$ 50 $\pm$ 20 $h^{-1}$ pc.

If we assume an axially symmetric disk of gas with a uniform structure
of temperature $\langle T_e\rangle$, free electron density $\langle
N_e\rangle$, and radius $R$, with a hole in the middle of radius $r$.
The optical depth at each point is
$$\tau_\nu \approx 0.08235 ~\nu^{-2.1} \langle T_e\rangle^{-1.35}
\langle N_e\rangle^2 L,\eqno(3)$$ where $L$ is the path length through
the ionized part of the disk to the radio emission in pc.  The values
in Figure 5, then, correspond to $0.08235 \langle T_e\rangle^{-1.35}
\langle N_e\rangle^2 L,$ which is roughly the emission measure
($\langle N_e\rangle^2 L$) times $\langle T_e\rangle^{-1.35}$ times a
constant.  For convenience of discussion we refer to the quantity
$[0.08235 \langle T_e\rangle^{-1.35} \langle N_e\rangle^2 L]$ as
$EMT^{-1.35}$.  When we average the maps in Figure 5 together we find
peak values of $EMT^{-1.35}$ of 10.6 to the north and 10.1 to the
south.  Reasonable values of $T_e$ and $N_e$ for the inner ionized
region of a gas disk of order 100 pc in radius in a radio galaxy
nucleus are $10^4$ K and 500 cm$^{-3}$, respectively, which yield a
path length of $L \approx 100$ pc.  We therefore find that entirely
reasonable physical parameters fit the observed free-free optical
depths.

Comparing our results to the H{\sc i} absorption results of Carilli et
al.\ (1998), we have
$${{\langle N_e\rangle^2 L_{e}} \over {\langle N_{\rm H{\sc i}}\rangle
L_{\rm H{\sc i}}}} \approx 1.3 \times 10^{-19} f {{\langle
T_e\rangle^{1.35}} \over {\langle T_s\rangle}}~{\rm cm}^{-3}~{\rm
K}^{-0.35},$$ where $f$ is the covering factor of the neutral gas and
$\langle T_s\rangle$ is the average spin temperature of the Hydrogen
atoms along the line of sight.  These results are reasonably
consistent, as can be demonstrated by the application of Saha equation
with some simplifying assumptions.  For argument's sake, we assume
$\langle T_e\rangle \approx \langle T_s\rangle = T$, $f$ = 1, and
$L_{\rm H{\sc i}} \approx 100 L_{e}$ (i.e. that the inner 1\% of the
disk is ionized).  Then,
$${{\langle N_e\rangle^2} \over {\langle N_{\rm H{\sc i}}\rangle}}
\approx 1.3 \times 10^{-17}~T^{0.35}~{\rm cm}^{-3}~{\rm K}^{-0.35}.$$
If we further approximate the gas as containing only Hydrogen, then
the Saha equation can be written as
$${{\langle N_e\rangle^2} \over {\langle N_{\rm H{\sc i}}\rangle}} =
2.4 \times 10^{15}~T^{3/2} \exp \left({-1.58\times 10^5~{\rm K} \over
T}\right)~{\rm cm}^{-3}~{\rm K}^{-3/2}.$$ Equating these two
expressions yields an average gas temperature of $\sim2000$ K.

\subsection{The Core}

The spectrum of the core region is already inverted at the observed
frequencies above 5 GHz.  The cause for the inverted spectrum of the
core, therefore, seems unrelated to the ionized disk of gas.  Since
the core is the region most likely to have large magnetic fields and
large densities of relativistic energies, we suspect that the inverted
spectrum of the core is due to synchrotron self-absorption.  With the
same 15.4-GHz data reported here, Taylor et al.\ (2000) infer an upper
limit to the magnetic field in the core, assuming synchrotron
self-absorption, of $7 \times 10^4$ Gauss.

\subsection{Probable Cause for Onset of Recent Activity}

The suggestion of a dense circumnuclear disk of gas in 0108+388 may
provide further insight into some of the unique characteristics of
this CSO.  We speculate that instabilities in such a disk could result
in periodic infall of gas, which will produce renewed apparent
activity.  This would simultaneously explain the extended emission 20
arcsec (78 $h^{-1}$ kpc) to the East detected by Baum et al.\ (1990)
and the young age of the compact structure inferred by Owsianik et
al.\ (1998) and Taylor et al.\ (2000).

\subsection{Implications for GPS Sources In General}

As the results presented here demonstrate, when a radio source's flux
density is produced by a number of components, absorption on large
scales is needed in order for the composite spectrum to have a sharp
turnover.  Synchrotron self-absorption depends on the internal
conditions of the source, and these conditions will vary from one part
of the source to another.  Such a large-scale absorption is therefore
unlikely to be due to this process.  On the other hand, we have also
found that the core of this particular GPS source is probably
synchrotron self-absorbed.  Its spectrum is already turned over at the
frequency where the free-free absorption becomes significant.  The
implication for other, and possibly all, GPS sources is that free-free
absorption is the likely cause for the high-frequency turnovers of the
composite spectrum, but synchrotron self-absorption may still be
important on smaller scales; the turnover of a weak core component
will not contribute significantly to the integrated flux densities at
low frequencies.

\section{CONCLUSIONS}

By extrapolating our high frequency maps, we obtained opacity maps of
0108+388 at two frequencies below the spectral turnover which strongly
suggest that free-free absorption, possibly by a disk of ionized gas,
is indeed the cause of the spectral turnover, and perhaps in other
CSOs with a GHz-peaked spectrum.  The spectra all across the map fit
free-free absorption well.  Reasonable values for temperature (10$^4$
K) and electron density (500 cm$^{-3}$) in a disk of order 100 pc in
size fit the observed optical depths.

The spectrum of the core, though, is inverted even at our highest
frequencies, probably because of synchrotron self-absorption.

The presence of a dense gas disk in 0108+388, which is also consistent
with the large column densities implied by the H{\sc i} absorption
measurements of Carilli et al.\ (1998), may be related to the recent
onset of activity in this object, as indicated by the age estimates by
Owsianik et al.\ (1998) and Taylor et al.\ (2000).  The accumulated
evidence thus strongly supports the conclusions that CSOs in general
are intrinsically young objects that evolve into classical kpc-scale
radio galaxies, and that the radio galaxy 0108+388, in particular, is
the youngest source known.

\acknowledgements 

This research was aided, in part, by funding from the Dudley
Observatory.  This research has also made use of data from the
University of Michigan Radio Astronomy Observatory which is supported
by funds from the University of Michigan.  We are also grateful for
the friendly assistance of Vivek Dhawan and the staff at the National
Radio Astronomy Observatory.



\begin{figure}
\phantom{mark}
\centerline{\psfig{figure=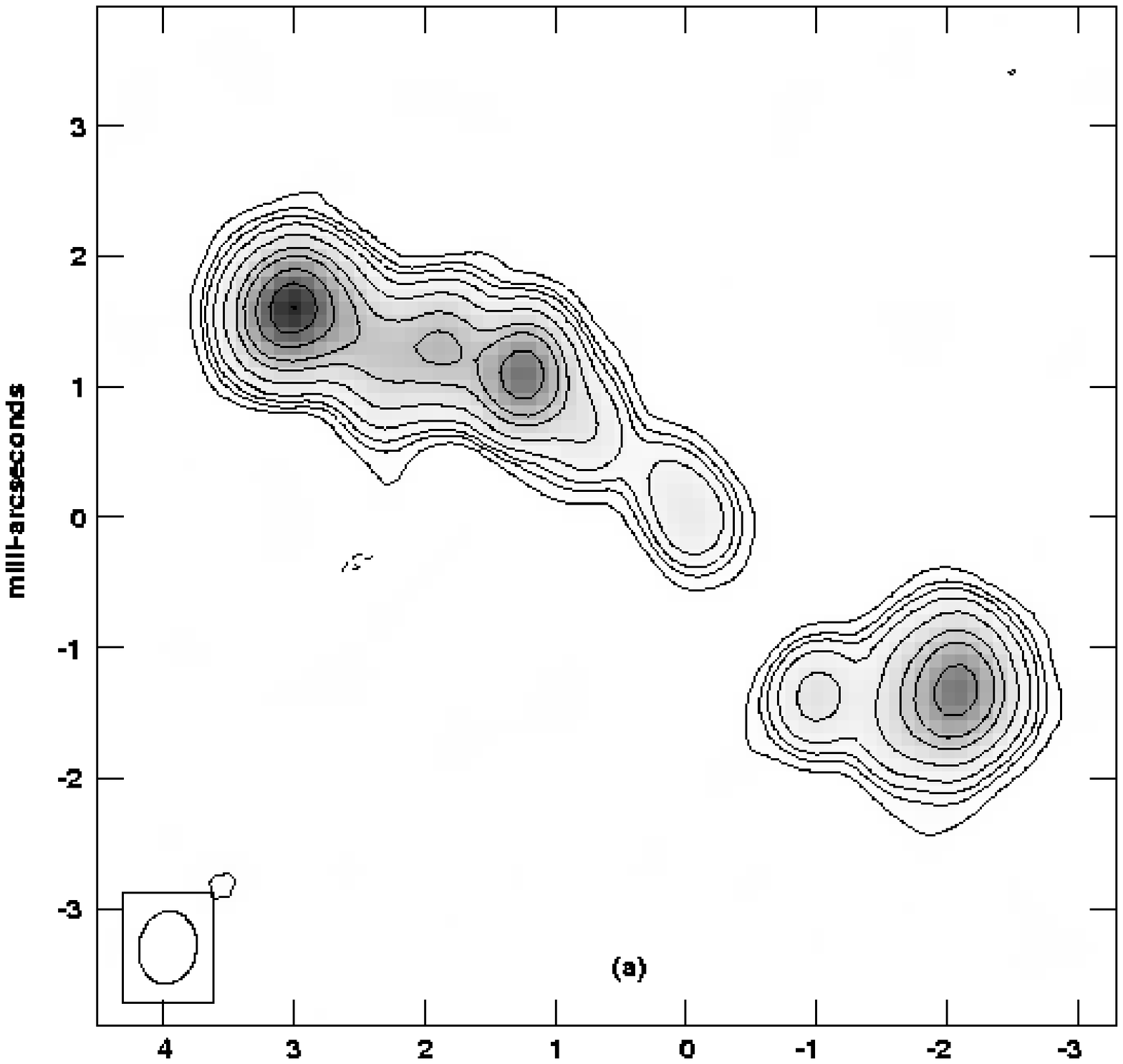,width=2.5in} 
            \psfig{figure=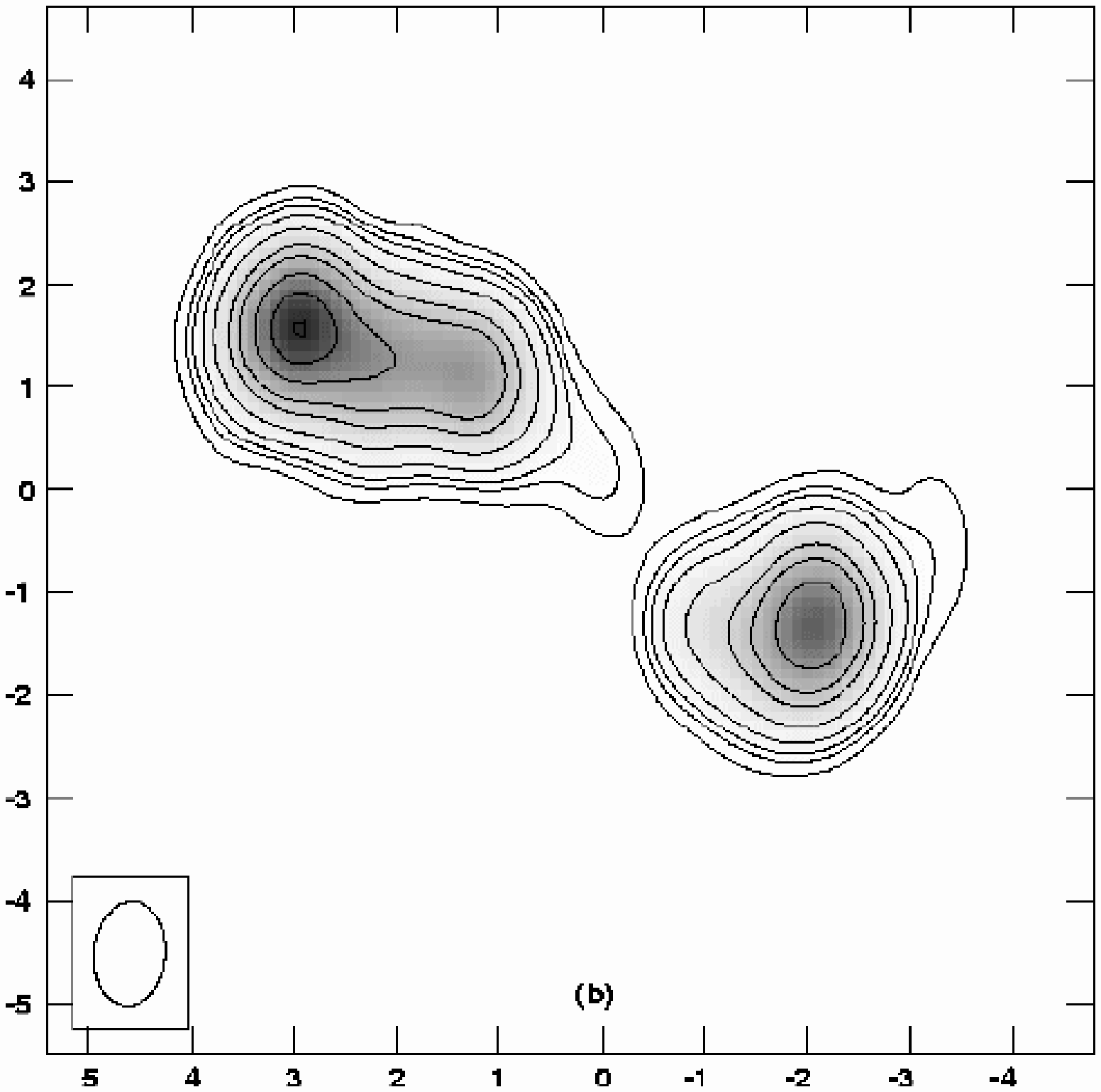,width=2.5in}}
\vspace{-0.75in}
\centerline{\psfig{figure=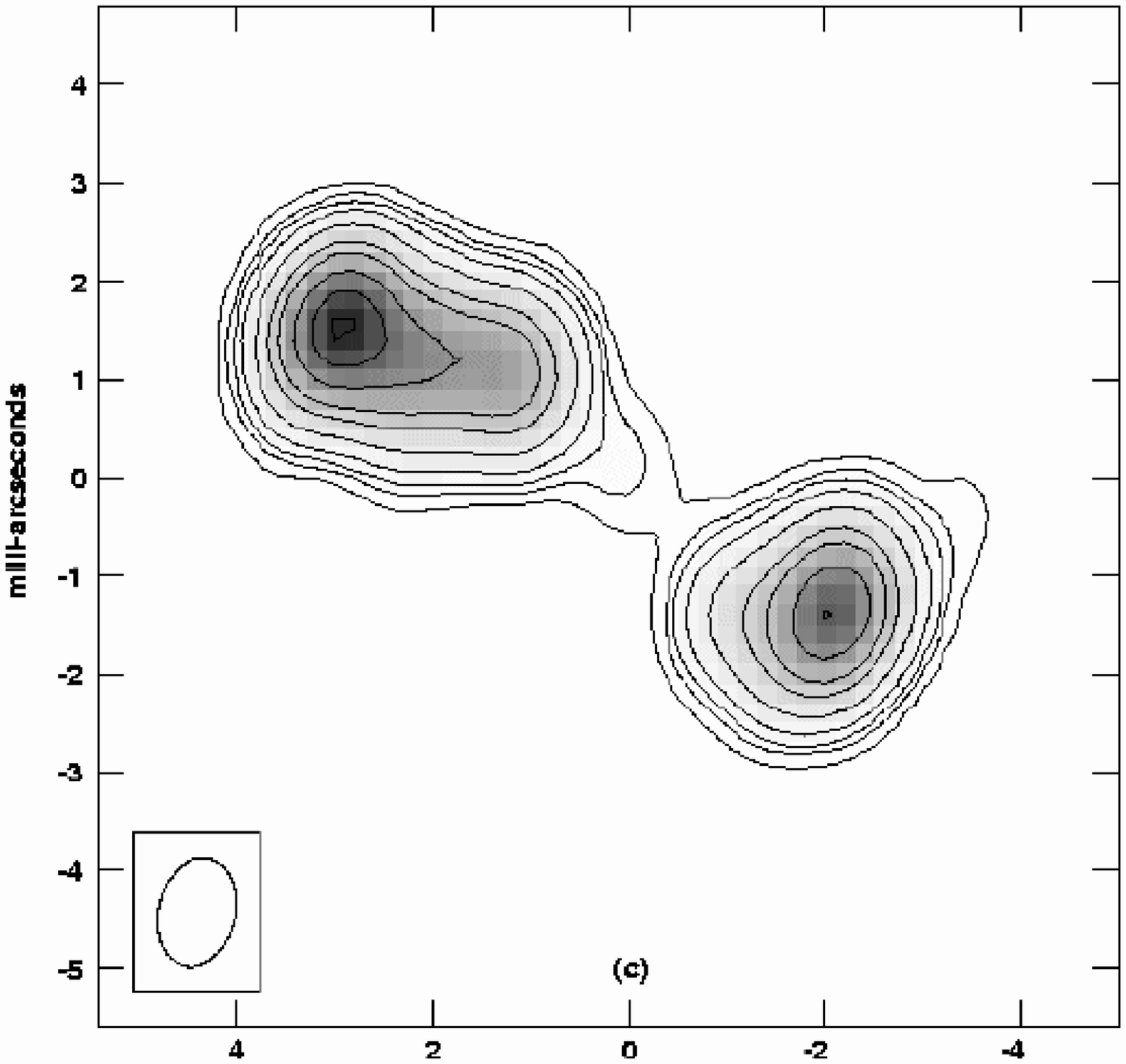,width=2.5in} 
            \psfig{figure=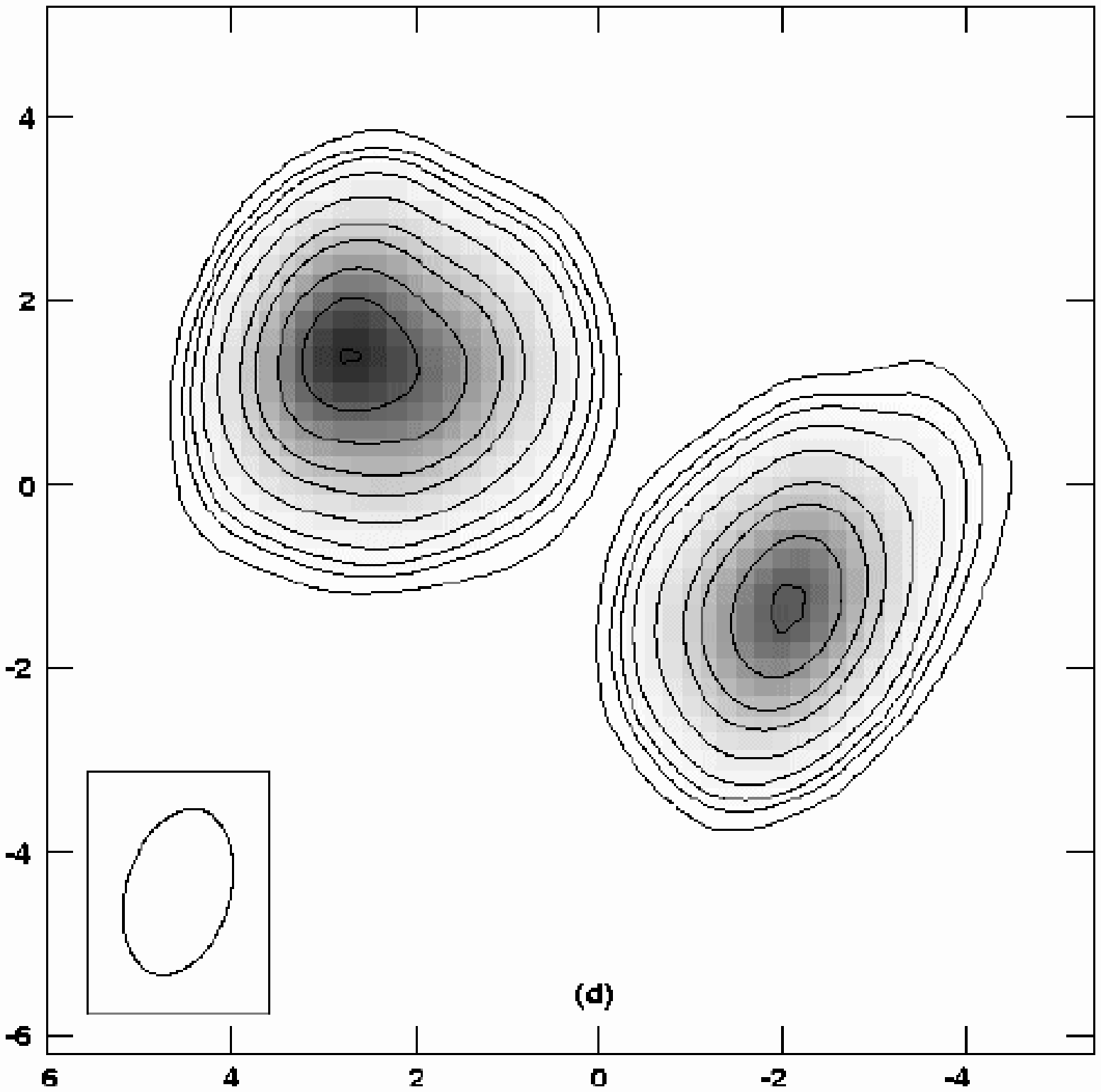,width=2.5in}}
\vspace{-0.75in} 
\centerline{\psfig{figure=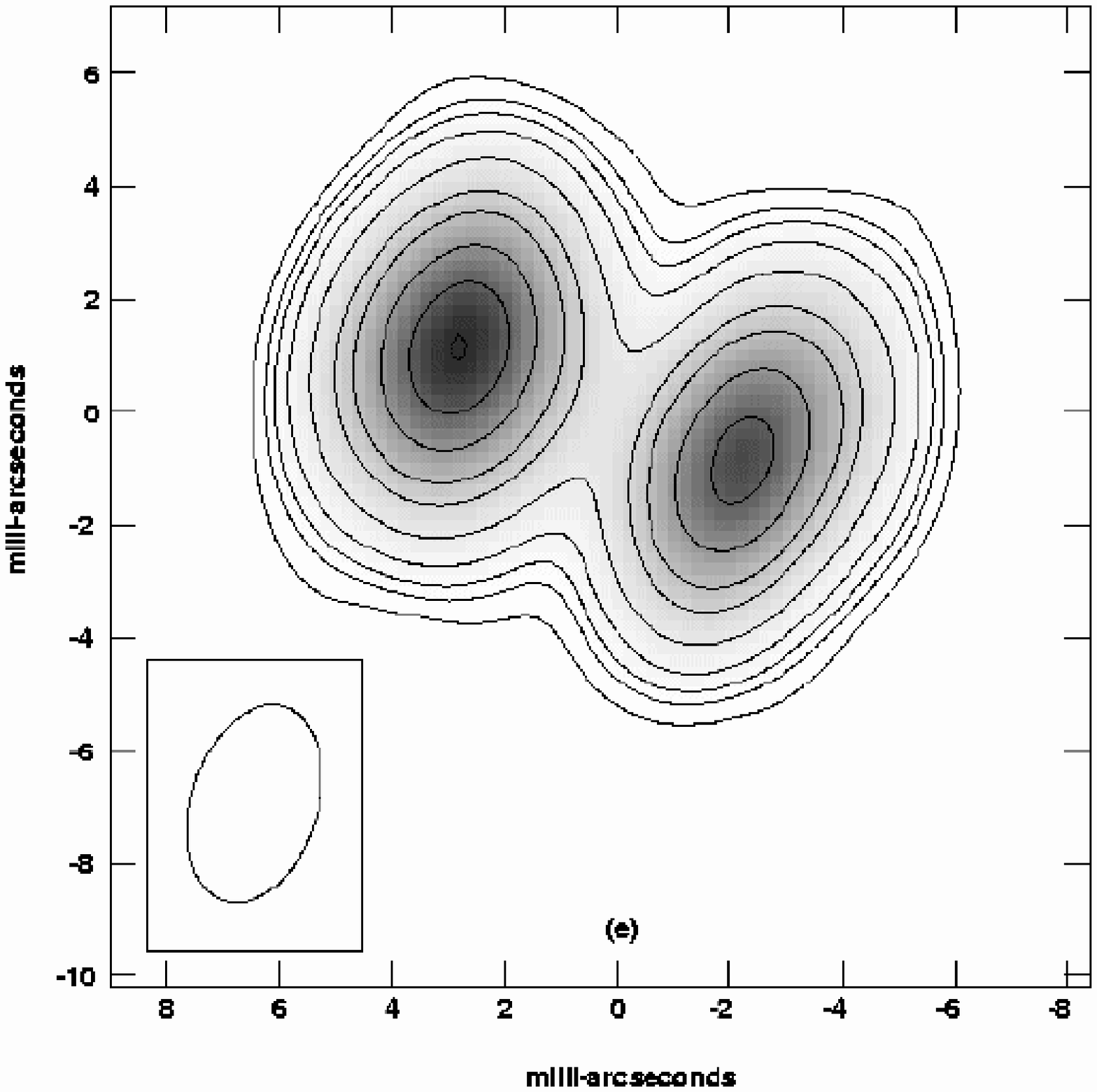,width=2.5in} 
            \psfig{figure=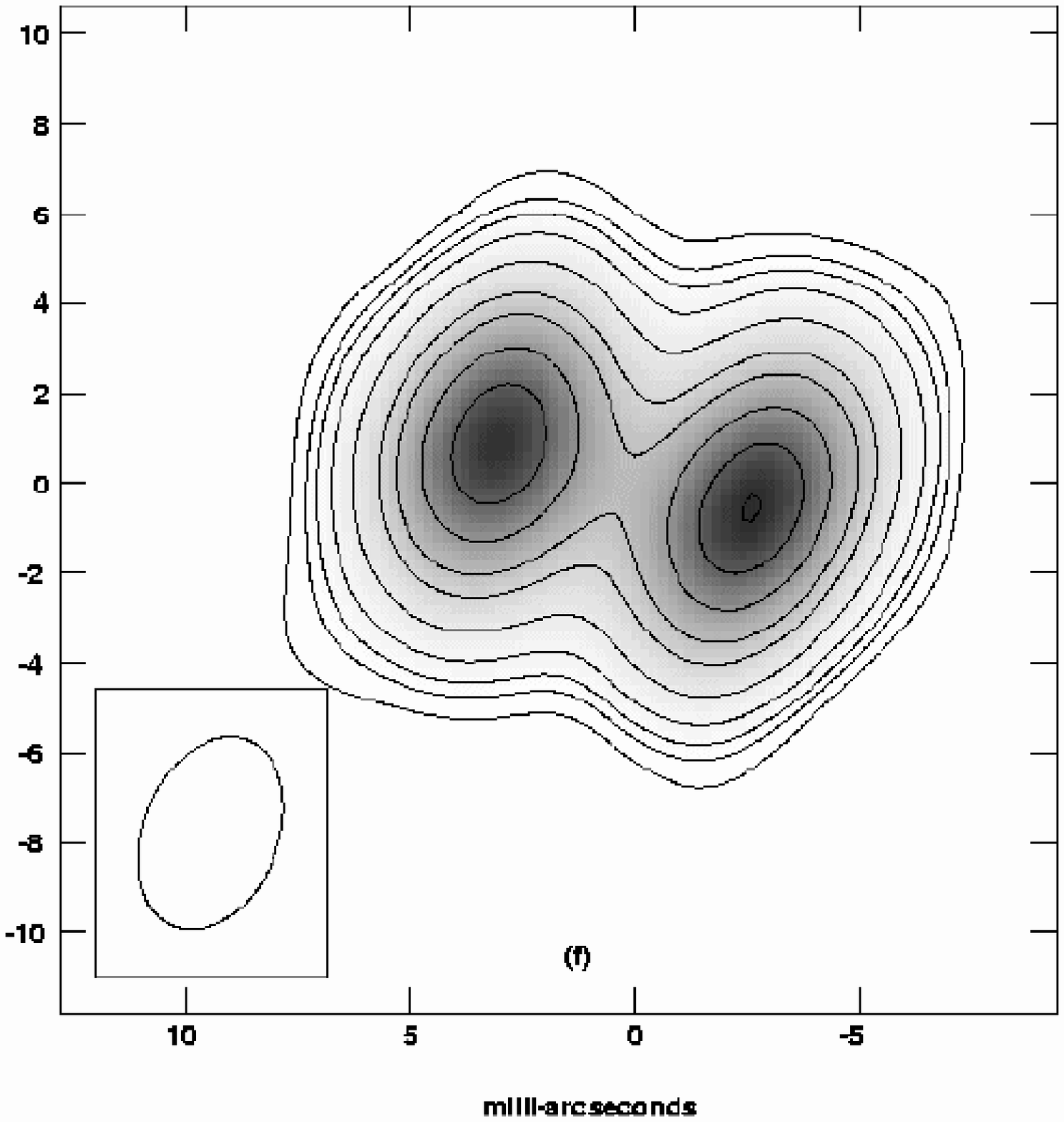,width=2.5in}}
\caption{Clean maps at six frequencies and two epochs:
(a) 15.4 GHz, 1997 Feb; (b) 8.4 GHz, 1997 Feb; (c) 8.4 GHz, 1996 Jun;
(d) 5.0 GHz, 1996 Jun; (e) 2.3 GHz, 1996 Jun; and (f) 1.7 GHz, 1996 Jun.  
The displayed contours represent $-$3, $-$2, $-$1, 1, 2,
3, 5, 10, 20, 30, 50, 75, and 99\% of the peak value.  
The peak values and rms noise levels are:
(a) 0.1040 \& 0.00033 Jy/beam; (b) 0.2228 \& 0.00020 Jy/beam;
(c) 0.2279 \& 0.00025 Jy/beam; (d) 0.3933 \& 0.00037 Jy/beam; 
(e) 0.3468 \& 0.00069 Jy/beam; and (f) 0.2217 \& 0.00069 Jy/beam.  
The FWHM of the clean beams (displayed in the lower left corners)
are:
(a) 0.56$\times$0.44 mas, PA=$-$8.6$^\circ$; 
(b) 1.01$\times$0.70 mas, PA=$-$5.0$^\circ$;
(c) 1.11$\times$0.78 mas, PA=$-$13.3$^\circ$; 
(d) 1.84$\times$1.15 mas, PA=$-$16.9$^\circ$;
(e) 3.61$\times$2.22 mas, PA=$-$16.7$^\circ$; 
(f) 4.46$\times$2.98 mas, PA=$-$21.2$^\circ$.
The maps are not all displayed to the same scale.}
\end{figure}

\begin{figure}
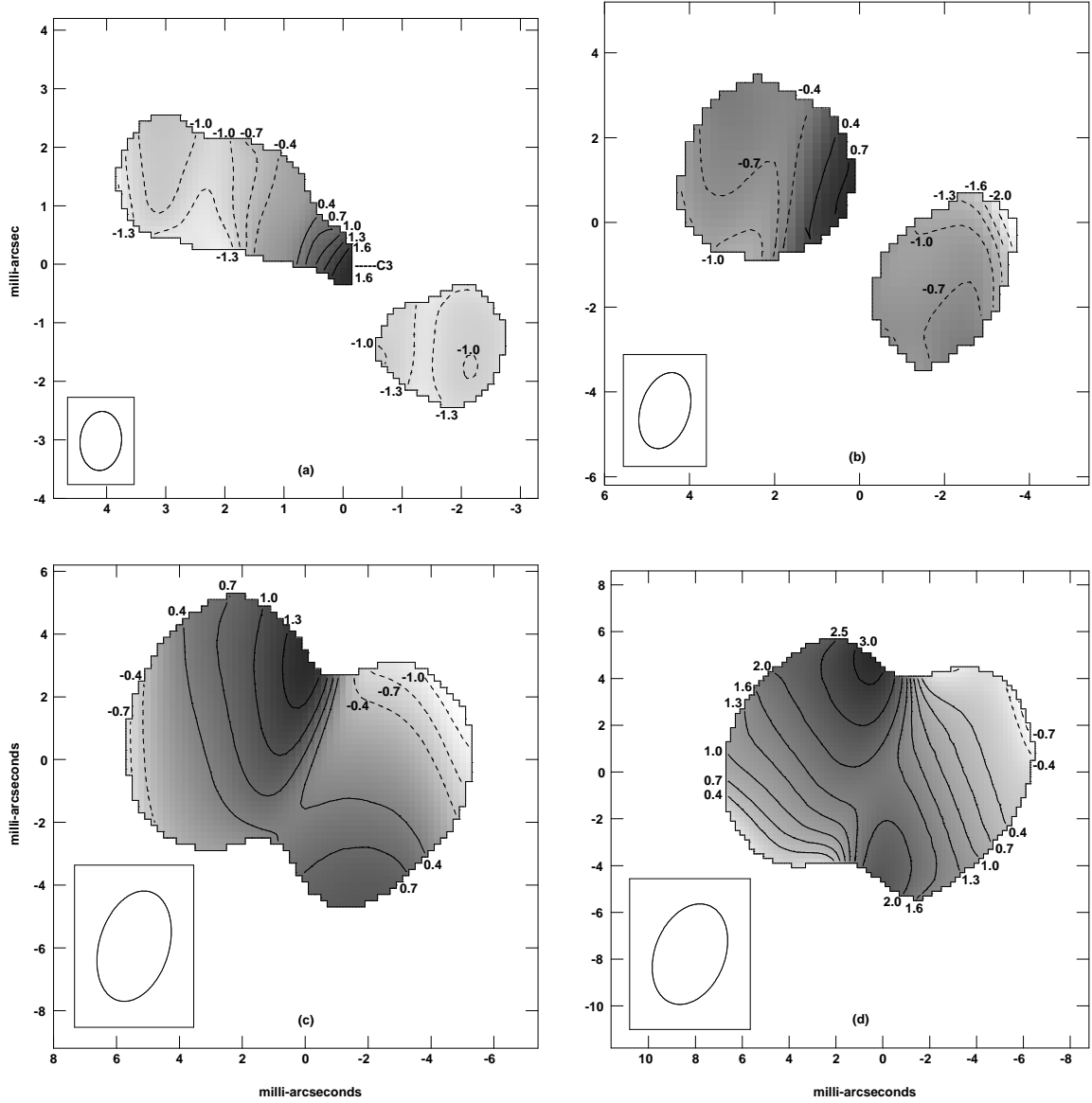

\centerline{\psfig{figure=f2a.ps,width=3.0in}
            \psfig{figure=f2b.ps,width=3.0in}}
\vspace{-0.5in} 
\centerline{\psfig{figure=f2c.ps,width=3.0in}
            \psfig{figure=f2d.ps,width=3.0in}}
\caption{Spectral index maps:
(a) between 15.4 and 8.4 GHz, 1997 Feb; (b) between 8.4 and 5.0 GHz, 1996 Jun;
(c) between 5.0 and 2.3 GHz, 1996 Jun; and 
(d) between 2.3 and 1.7 GHz, 1996 Jun.  
The labels of the contours indicate spectral index values, $\alpha$, where
$F_\nu \propto \nu^\alpha$.  Solid 
contours represent positive spectral indices and dashed contours represent
negative spectral indices.  The grey scale is darker towards larger 
spectral indices.  The uncertainties in the spectral indices are 
listed in Table 1.
The input clean maps were both convolved with clean beams, displayed
in the lower left corners, equal to that of the lower frequency map in
Figure 1.
The maps are not all displayed to the same scale.}
\end{figure}


\begin{figure}
\centerline{\psfig{figure=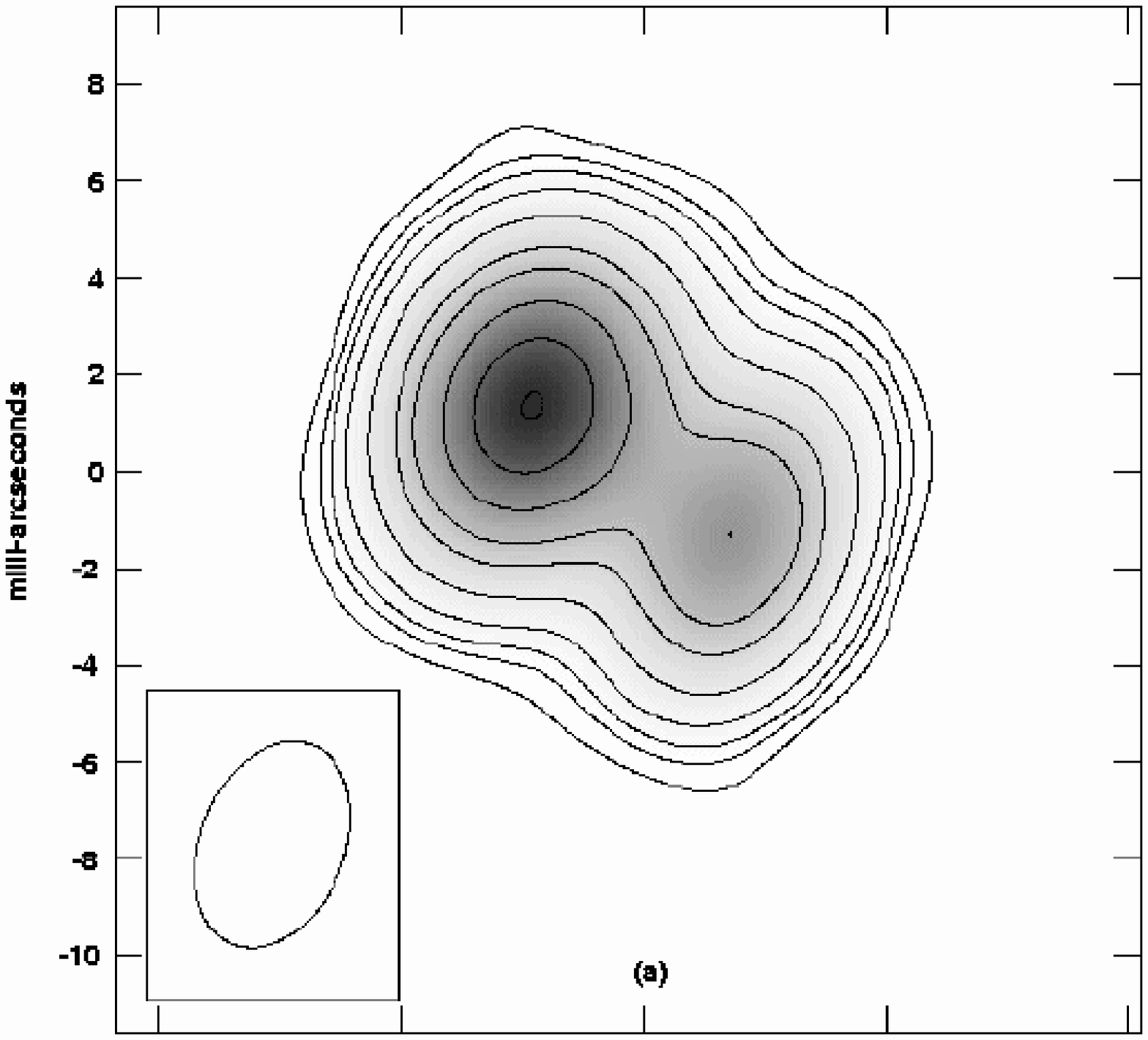,width=2.5in}
            \psfig{figure=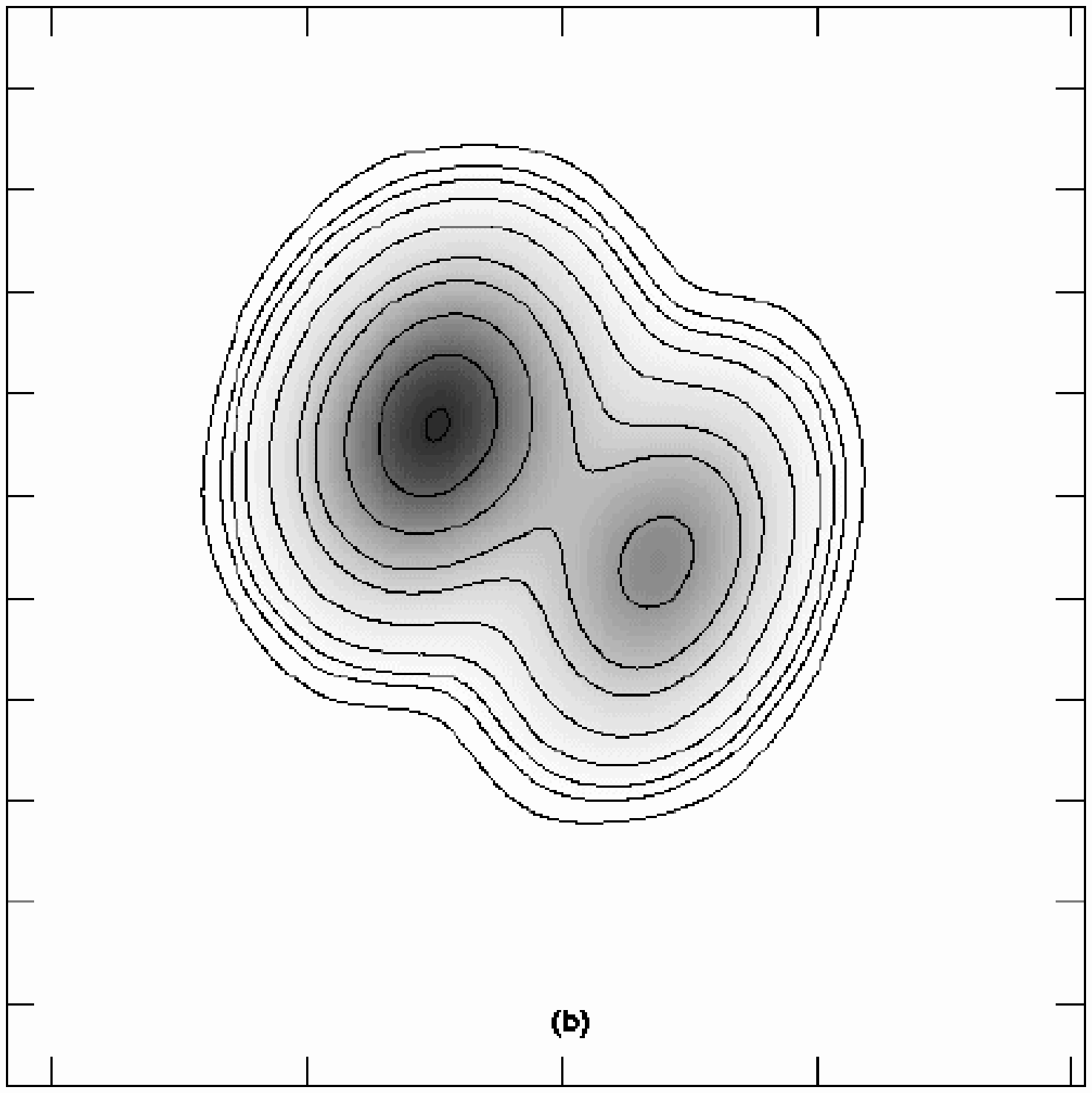,width=2.5in}}
\vspace{-0.75in}
\centerline{\psfig{figure=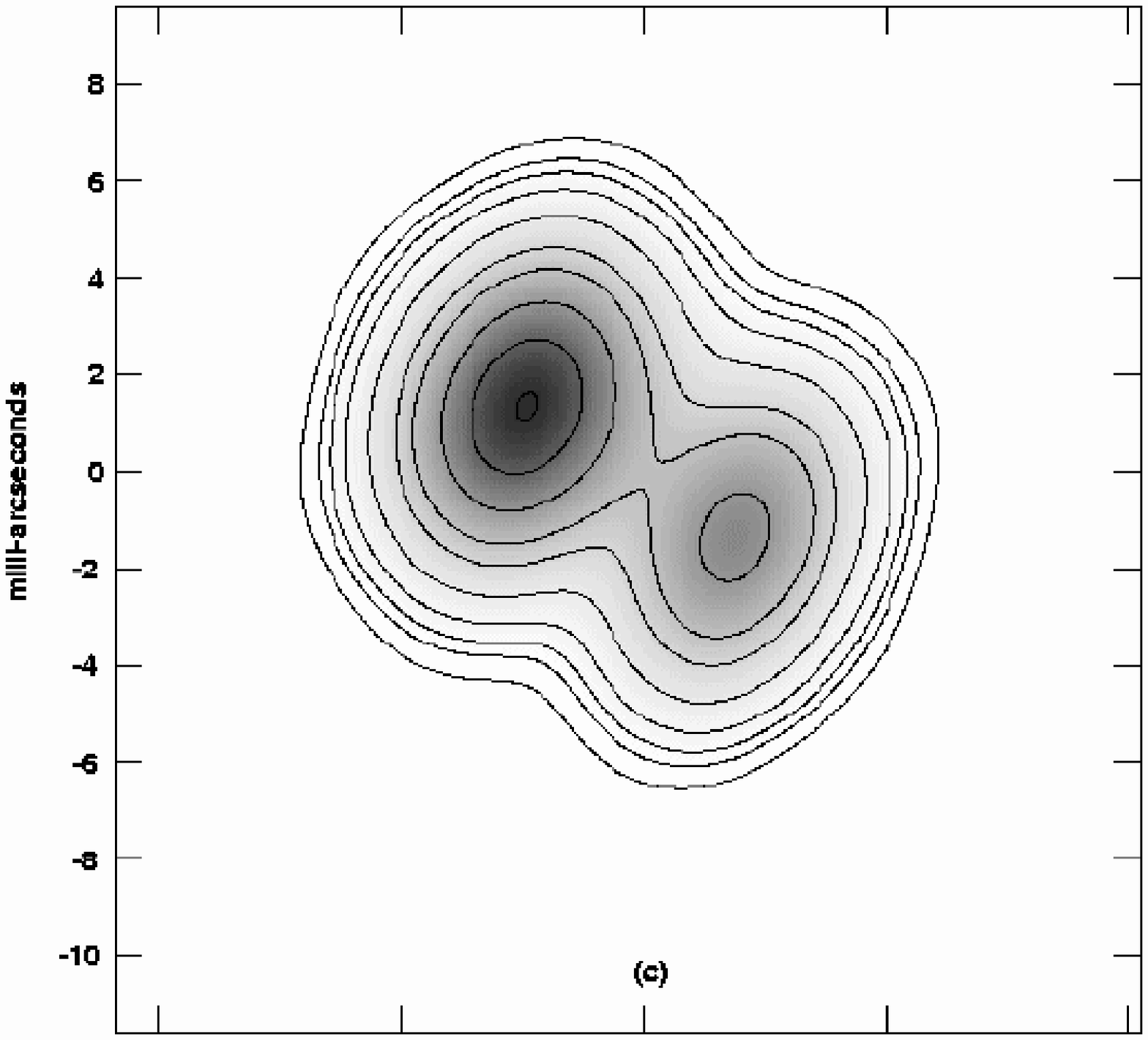,width=2.5in}
            \psfig{figure=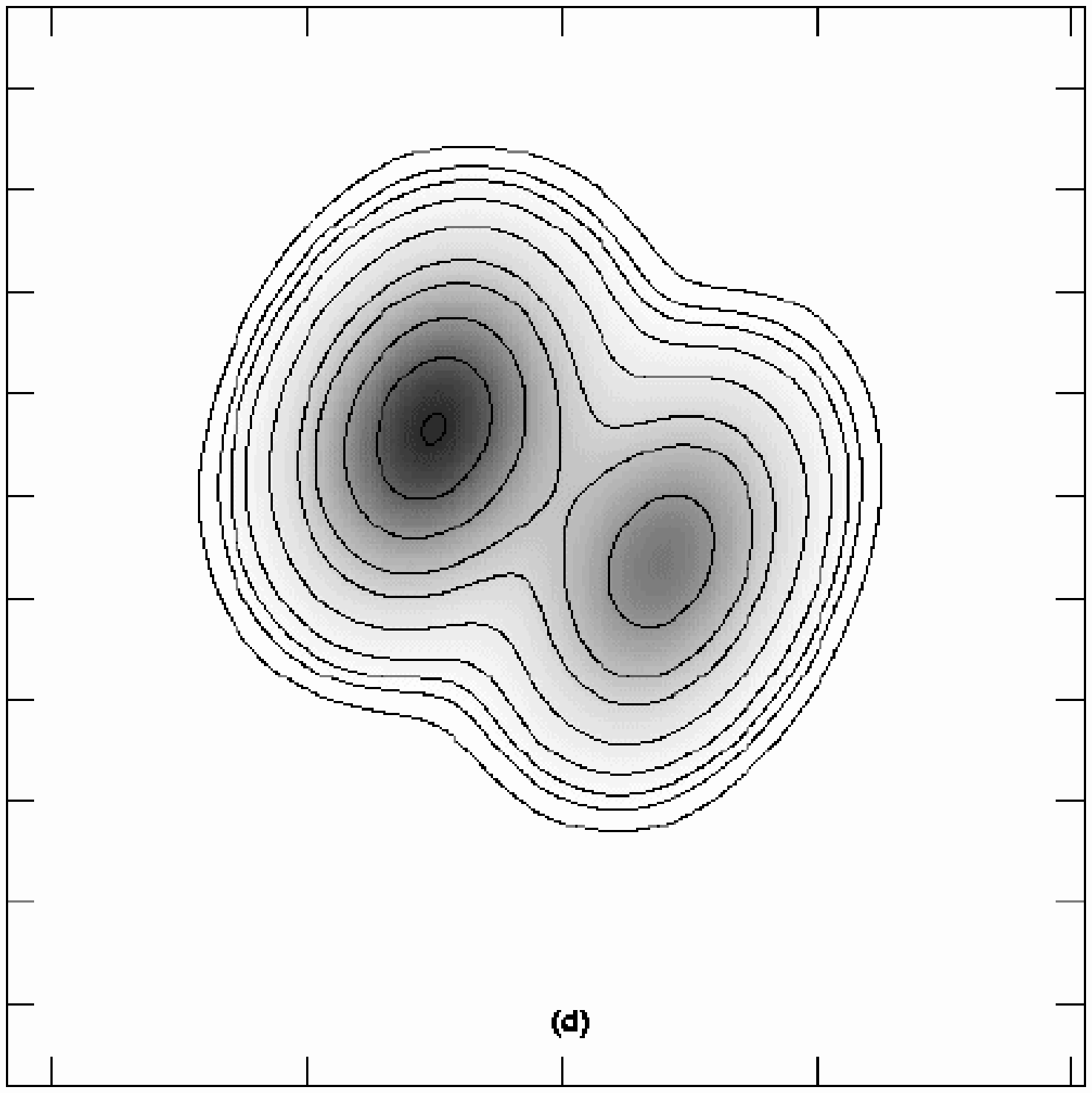,width=2.5in}}
\vspace{-0.75in}
\centerline{\psfig{figure=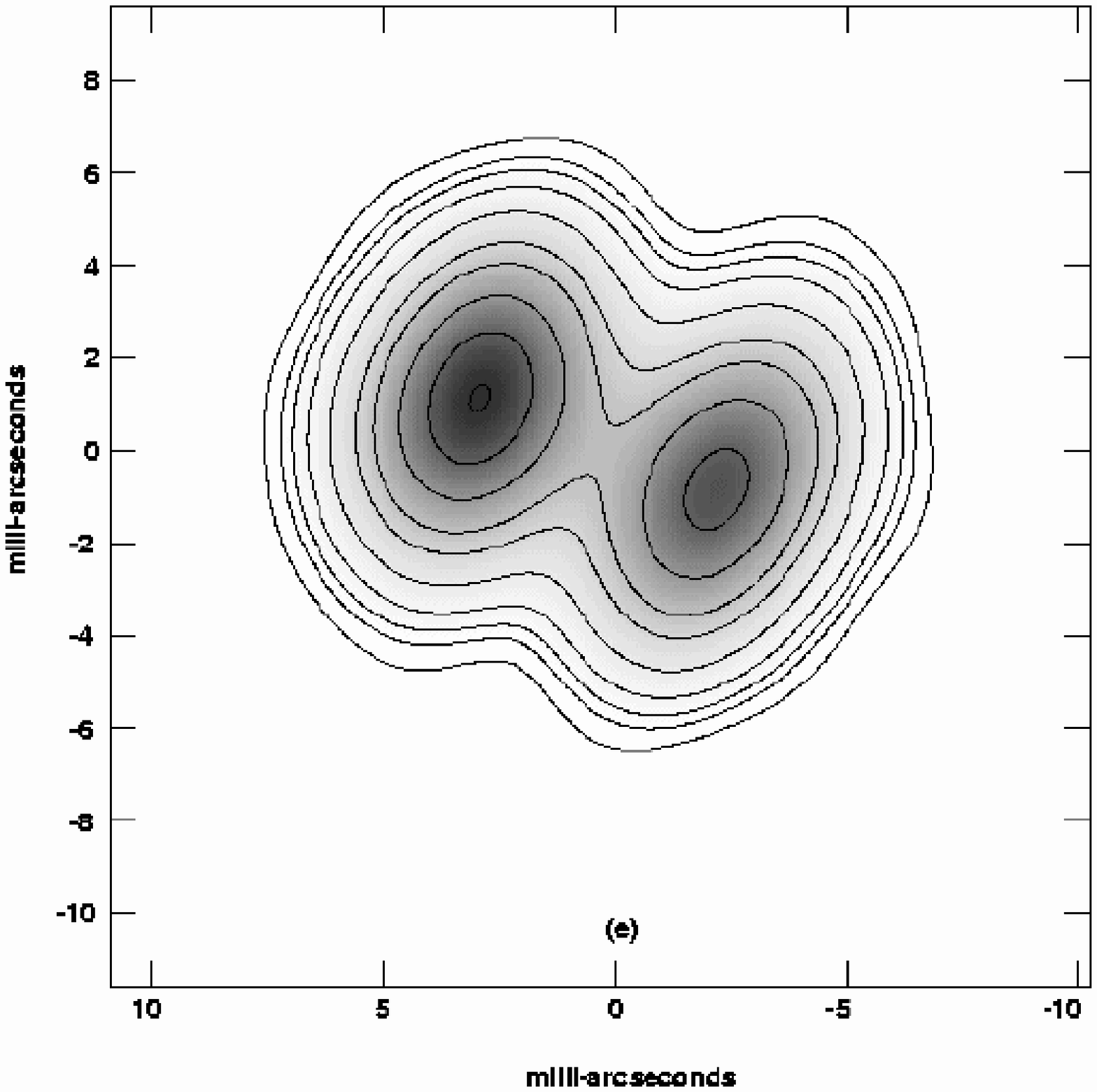,width=2.5in}
            \psfig{figure=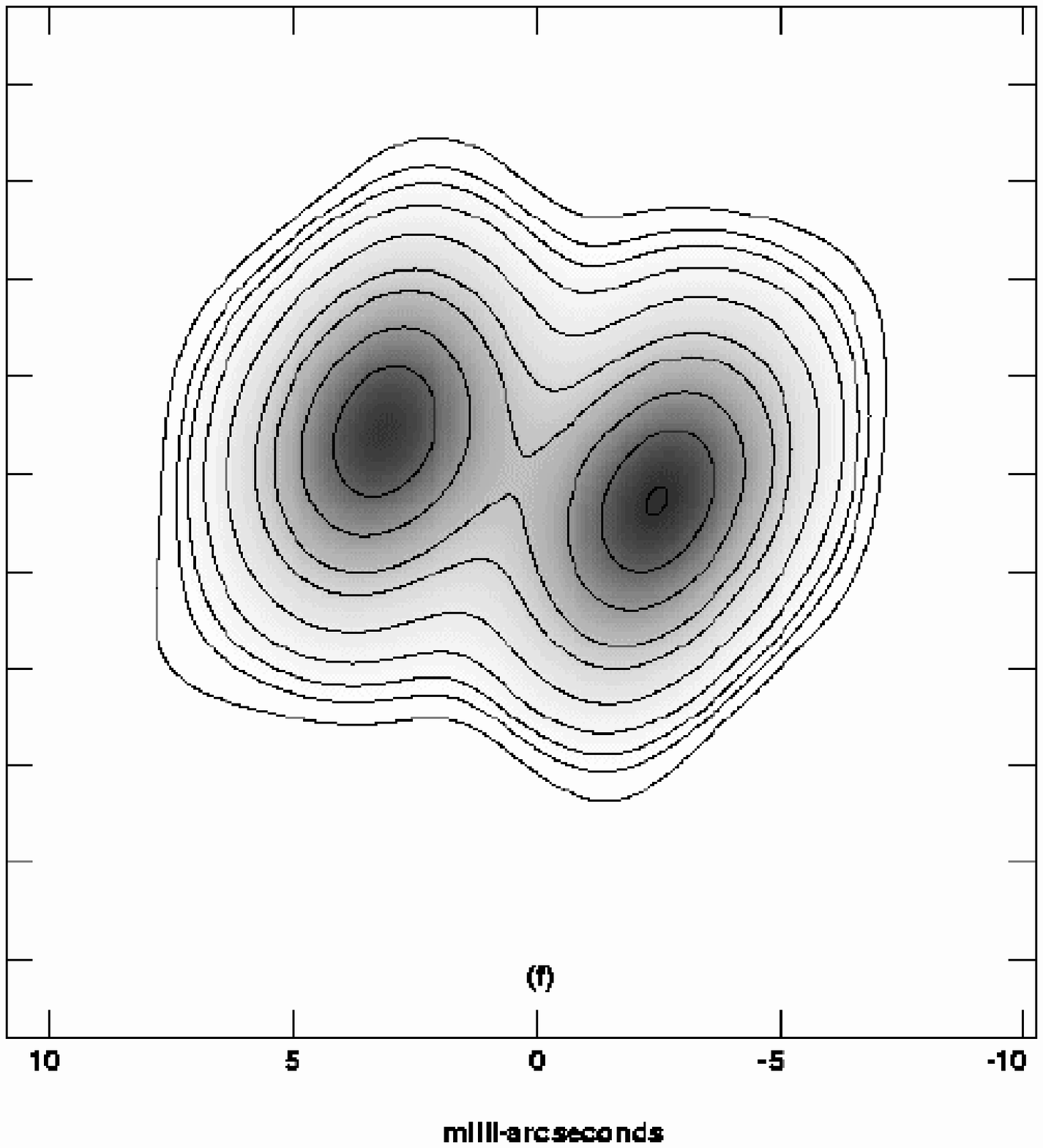,width=2.5in}}
\caption{Clean maps made from truncated data sets all with the same
minimum and maximum $u$-$v$ distances, and convolved with the same
clean beam: (a) 15 GHz, 1997 Feb; (b) 8.4 GHz, 1997 Feb; (c) 8.4 GHz,
1996 Jun; (d) 5.0 GHz, 1996 Jun; (e) 2.3 GHz, 1996 Jun; and (f) 1.7
GHz, 1996 Jun.  The displayed contours represent $-$3, $-$2, $-$1, 1,
2, 3, 5, 10, 20, 30, 50, 75, and 99\% of the peak value.  The peak
values and rms noise levels are: (a) 0.2713 \& 0.00069 Jy/beam; (b)
0.4738 \& 0.00054 Jy/beam; (c) 0.4617 \& 0.00028 Jy/beam; (d) 0.6218
\& 0.00036 Jy/beam; (e) 0.3788 \& 0.00054 Jy/beam; and (f) 0.2212 \&
0.00063 Jy/beam.  The FWHM of the clean beam, which is the same for
all images and is displayed in the lower left corner of figure (a), is
4.44$\times$2.97 mas at position angle $-$21.5$^\circ$.  The maps are
displayed to the same scale.}
\end{figure}


\begin{figure}
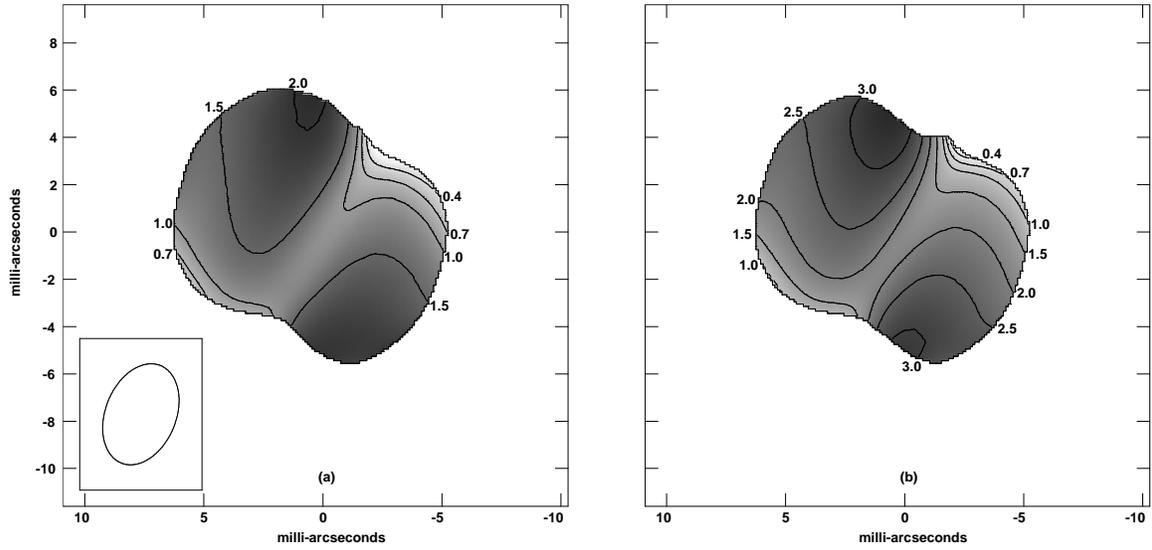

\centerline{\psfig{figure=f4a.ps,width=3.0in}
            \psfig{figure=f4b.ps,width=3.0in}}
\caption{Opacity maps, made from data sets edited
to yield the same resolution, at
(a) 2.3 GHz and (b) 1.7 GHz. 
The labels of the contours indicate optical depth values with the grey scale 
becoming darker towards larger optical depths.  
The uncertainties in the optical depths are listed in Table 2.}
\end{figure}


\begin{figure}
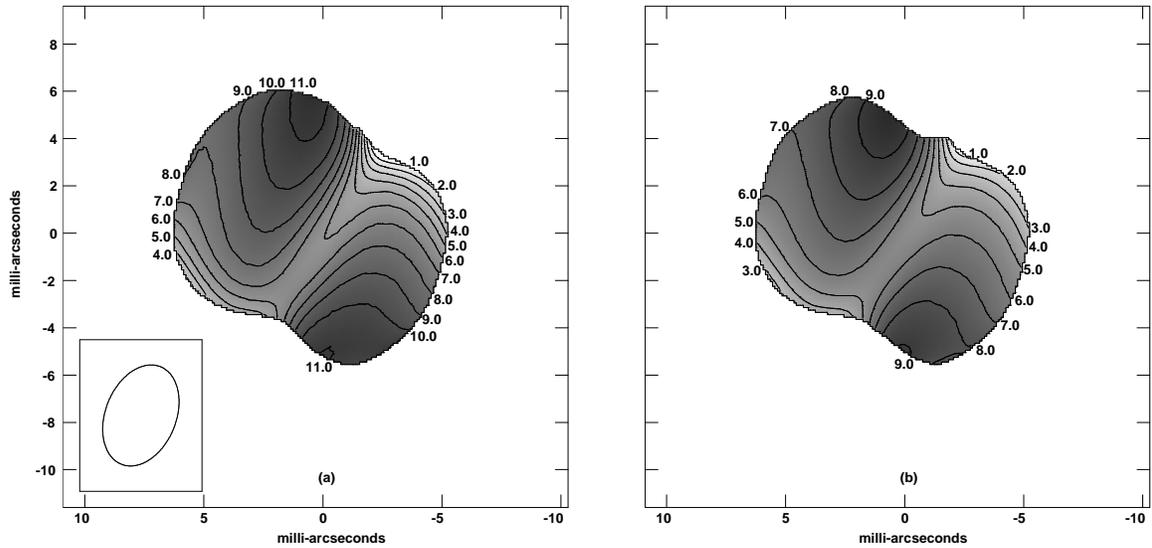

\centerline{\psfig{figure=f5a.ps,width=3.0in}
            \psfig{figure=f5b.ps,width=3.0in}}
\caption{Maps of the quantity 
$0.08235 \int T_e^{-1.35} N_e^2 dl$ at (a) 2.3 and (b) 1.7 GHz.
The uncertainties are listed in Table 2.}
\end{figure}


\begin{figure}
\centerline{\psfig{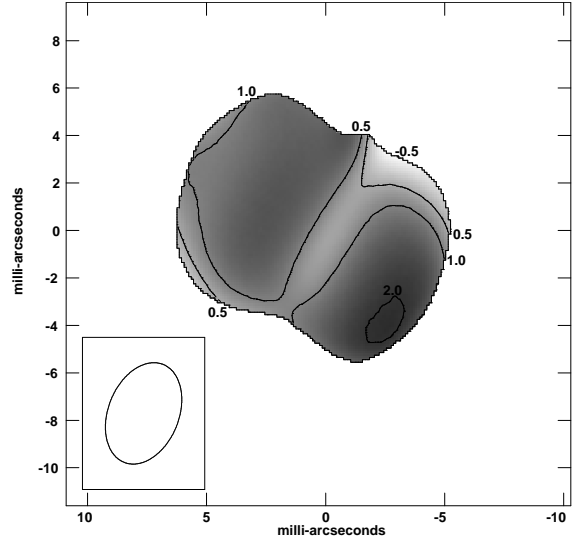}}
\caption{The difference between the
maps in Figures 5(a) and 5(b) divided by the calculated uncertainties in this
difference.  The contours represent difference values of
$-$0.5, 0.5, 1, and 2 times the uncertainty.}
\end{figure}

\bigskip

\begin{figure}
\centerline{\psfig{figure=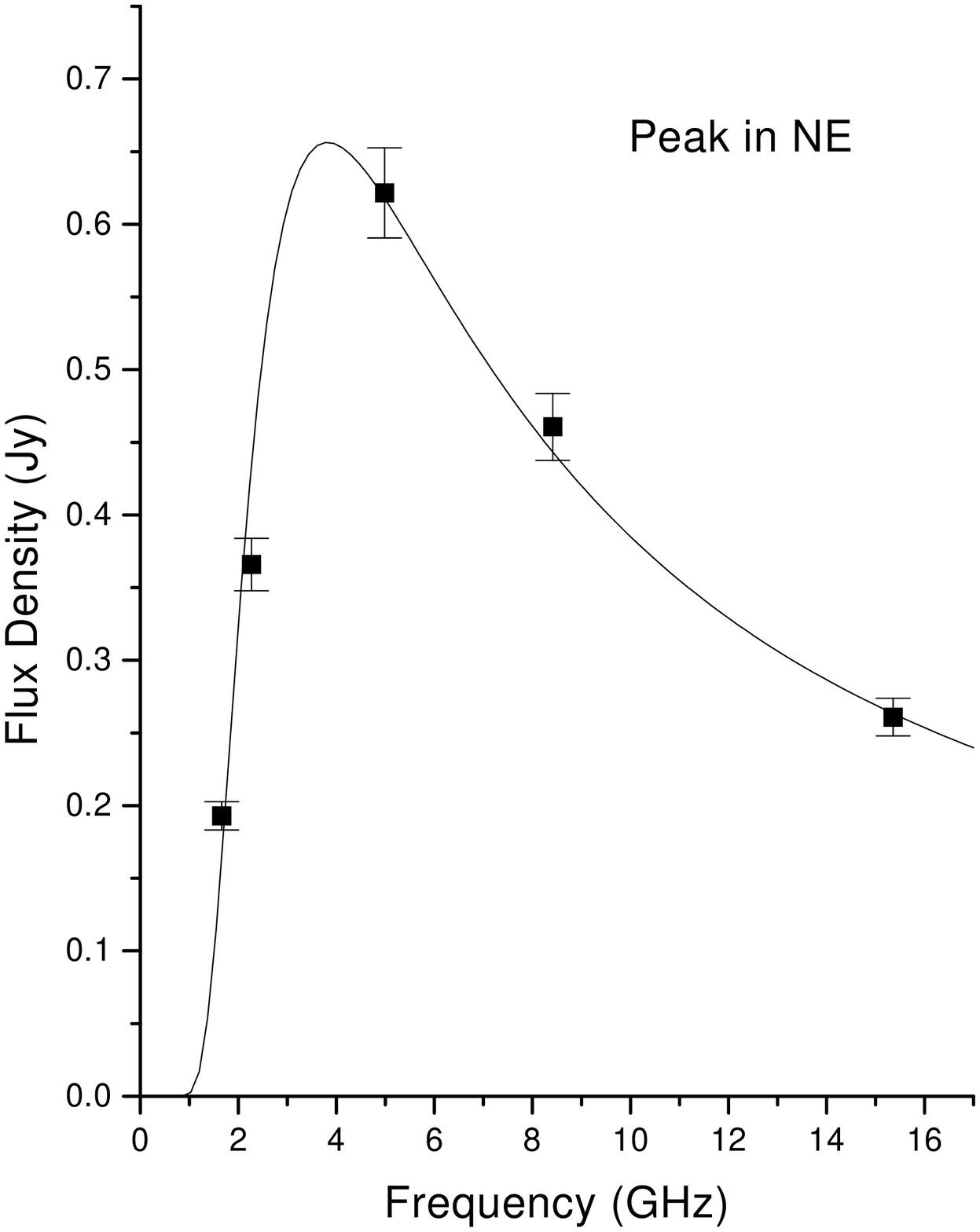,width=3.0in}
            \psfig{figure=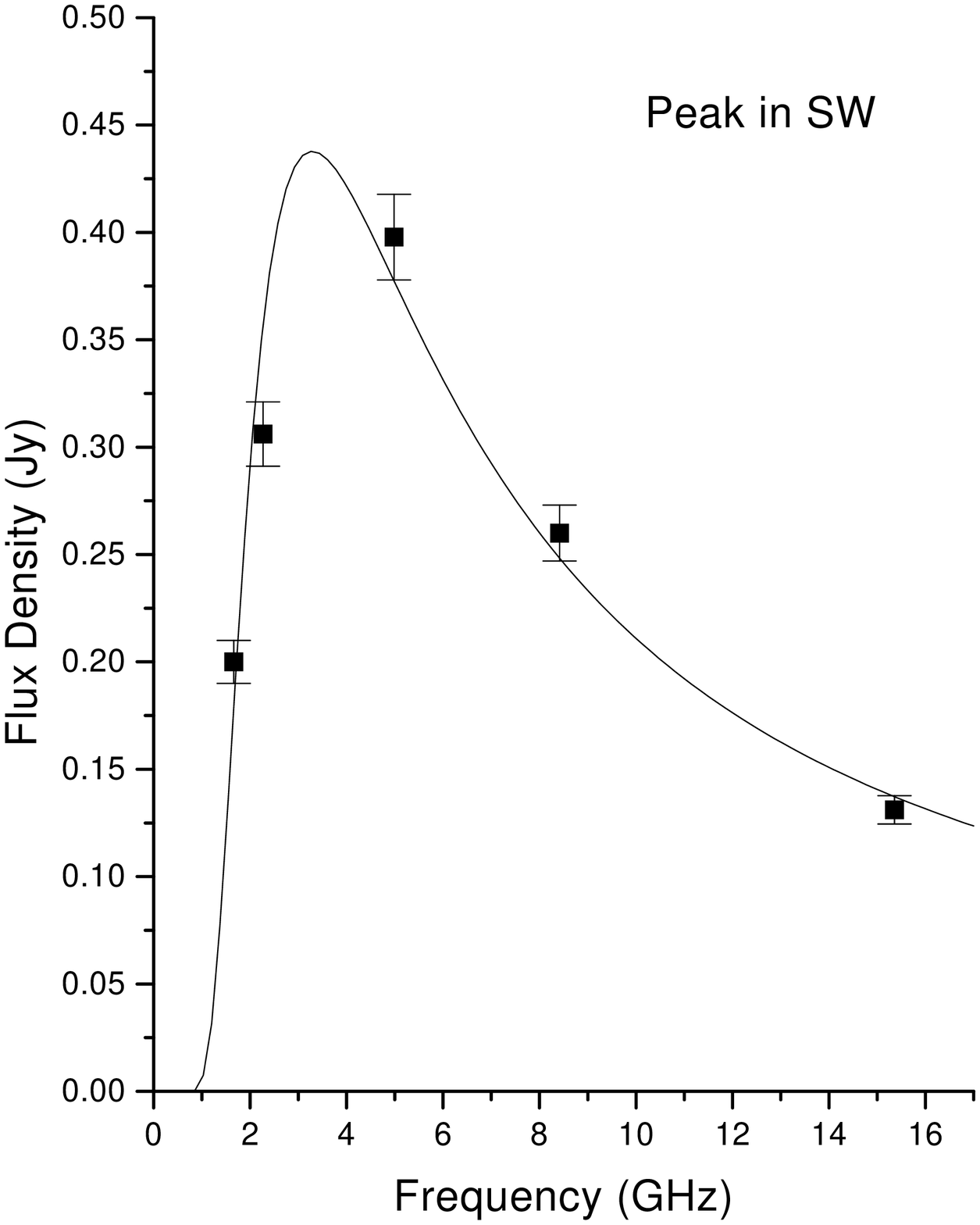,width=3.0in}}
\caption{Spectra at the peaks in emission at 5.0 GHz.  
The error bars represent the total
uncertainties in the flux densities.  Overlaid as 
continuous lines are the  best-fit curves of free-free absorption by 
single components in the foreground of a transparent synchrotron source.}
\end{figure}


\begin{figure}
\centerline{\psfig{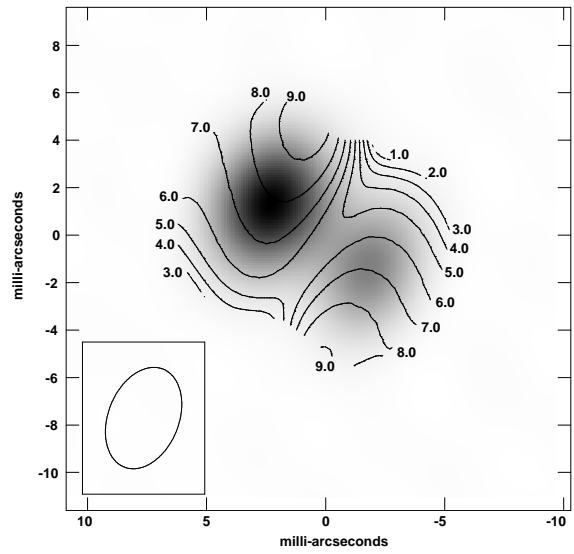}}
\caption{Contours from Figure 5(b) overlaid on a grey-scale plot of 
the emission at 15.4 GHz, as displayed in Figure 3(a).}
\end{figure}

\clearpage

\begin{table}[htb]
\begin{center}
\caption{Uncertainties in spectral index maps due to uncertainties in
the observed flux densities.  Uncertainties in observed flux densities
result both from possible errors in the amplitude calibration and from
map noise.  The uncertainties in $\alpha$ due to amplitude calibration
errors, listed in column 2, are constant across each map.  The
uncertainties in $\alpha$ due to map noise varies and depends on the
ratio of the observed flux density to the map rms.  Since the spectral
index maps were truncated where the map signals were $< 10\sigma$, the
maximum uncertainty due to map noise, which occurs at the edges of the
source, are calculated and are listed in column 3.}
\bigskip
\begin{tabular}{ccc}
\hline
Frequencies (GHz) & $\Delta\alpha$ due to amp. cal. errors & 
Max. $\Delta\alpha$ due to map noise  \\
\hline
15.359-8.417  &  0.12~ & 0.24 \\
~8.417-4.983  &  0.13~ & 0.26 \\
~4.983-2.267  &  0.090 & 0.18 \\
~2.267-1.663  &  0.23~ & 0.46 \\
\hline
\end{tabular}
\end{center}
\end{table}

\medskip

\begin{table}[htb]
\begin{center}
\caption{Uncertainties in optical depths and in $EMT^{-1.35}$ based on
5\% uncertainty in amplitude calibration of the $u$-$v$ data.}
\bigskip
\begin{tabular}{cll}
\hline
Frequency (GHz) & $\Delta\tau$ & $\Delta EMT^{-1.35}$ \\ 
\hline
1.663  & 0.21 &  0.60 \\
2.267  & 0.17 &  0.96 \\
\hline
\end{tabular}
\end{center}
\end{table}

\medskip

\begin{table}[htb]
\begin{center}
\caption{Best fit values of the parameters in a simple free-free absorption 
model to the observed flux densities at the peaks in the 5.0-GHz emission.
The parameters $F_o$, $\alpha$, and $EMT$ are given in the model by
$F_\nu = F_o (\nu / 1.5359 {\rm GHz})^\alpha \exp(-EMT/\nu^{2.1}).$ 
The parameter $F_o$ was constrained (see text) while the other two
parameters were allowed to vary freely.}
\bigskip
\begin{tabular}{lcc}
\hline
Parameters & NE peak & SW peak \\
\hline
$F_o$ & 0.27 $\pm$ 0.03 & 0.14 $\pm$ 0.015 \\
$\alpha$ & $-$0.97 $\pm$ 0.06  &  $-$1.07 $\pm$ 0.07 \\
$EMT$ & 7.7 $\pm$ 0.8 & 6.2 $\pm$ 0.6 \\
$\chi_{\nu}^2$ & 7.5 & 6.1 \\
\hline
\end{tabular}
\end{center}
\end{table}

\end{document}